\documentclass[%
 reprint,
superscriptaddress,
 amsmath,amssymb,
 aps,
 pre,
]{revtex4-1}

\usepackage{graphicx}
\usepackage{dcolumn}
\usepackage{bm}
\usepackage[colorlinks= true, linkcolor=blue, citecolor=blue, urlcolor=blue]{hyperref}
\usepackage{lipsum}
\usepackage{bbold}
\usepackage{mathtools}
\usepackage[normalem]{ulem}
\usepackage{float}
\usepackage{adjustbox}
\usepackage{cancel}

\def\bea{\begin{eqnarray}}
\def\eea{\end{eqnarray}}

\usepackage{xcolor}

\newcommand{\eref}[1]{Eq.~(\ref{#1})}
\newcommand{\fref}[1]{Fig.~\ref{#1}} 

\begin{document}

\title{When proofreading improves both speed and accuracy}
\author{Arup Biswas}
\affiliation{The Institute of Mathematical Sciences, CIT Campus, Taramani, Chennai 600113, India }
\affiliation{ Homi Bhabha National Institute, Training School Complex, Anushakti Nagar, Mumbai 400094, India}
\affiliation{School of Engineering and Applied Sciences, Harvard University, Cambridge, MA 02138, USA}
\author{L. Mahadevan}
\email{\textcolor{black}{Contact author:} lmahadev@g.harvard.edu}
\affiliation{School of Engineering and Applied Sciences, Harvard University, Cambridge, MA 02138, USA}
\affiliation{Departments of Physics and Organismic and Evolutionary Biology, Harvard University, Cambridge, MA
02138, USA }


\begin{abstract} 
Proofreading is generally thought to improve accuracy at the expense of speed. We show that this trade-off can be reversed in stochastic processes with long-lived stalled states. Using a non-Markovian renewal framework, we derive exact expressions for the error rate and completion time under proofreading for arbitrary stall-time distributions. Our analysis reveals that fluctuations in stall durations, rather than their mean alone, determine whether proofreading can simultaneously increase speed and accuracy. In the limit of strong stalling, this regime emerges when the coefficient of variation of the stall time exceeds a threshold set by the intrinsic error rate. These results provide a general criterion for proofreading in systems ranging from self-assembly and polymer replication to immune recognition and other nonequilibrium information-processing systems. 
\end{abstract}

\pacs{Valid PACS appear here}
\begin{titlepage}
\maketitle
\end{titlepage}

\emph{\textbf{Introduction.}---}  Many biological and physical processes must operate rapidly while maintaining high fidelity. Examples range from polymer replication and transcription \cite{kunkel2005dna,poulton2019nonequilibrium,thomas1998transcriptional,zaher2009fidelity,voliotis2009backtracking,murugan2012speed} to self-assembly of viral capsids, microtubules, crystals, and nanostructures \cite{hagan2011mechanisms,dogterom1993physical,de2003principles,whitelam2016minimal,wang2016peptide}, as well as immune recognition by T cells \cite{mckeithan1995kinetic,goldstein2004mathematical}. In such systems, errors can compromise function, motivating proofreading mechanisms that reject incorrect states. Kinetic proofreading schemes, introduced more than half a century ago \cite{hopfield1974kinetic,ninio1975kinetic}, are known to enhance fidelity through repeated resets that consume energy but are generally expected to slow the process.

Recent work has shown that this speed–accuracy trade-off can be reversed \cite{ravasio2026evolution,ravasio2026conditioning}, particularly when incorrect states generate long-lived stalls or kinetically trapped configurations \cite{esteban1993fidelity,thomas1998transcriptional,joazeiro2019mechanisms,johnston2001rna}. By removing such states, proofreading can both reduce errors and rescue trajectories that would otherwise remain delayed. However, the arguments for when this can happen rely on the relative role of the mean stall-time \cite{ravasio2026evolution,ravasio2026conditioning}, leaving the effect of stall-time fluctuations largely unexplored.

Here we formulate proofreading as a non-Markovian renewal process with resetting, building on ideas from stochastic resetting and first-passage theory \cite{evans_diffusion_2011,evans2011diffusion,evans_stochastic_2020}. This framework naturally accommodates arbitrary proofreading and stall-time distributions \cite{bel2010simplicity,munsky2009specificity}, while still allowing for analytic tractability and see how stall-time fluctuations can be a fundamental determinant of proofreading efficiency. 

\begin{figure}
    \centering
    \includegraphics[width=8.4cm]{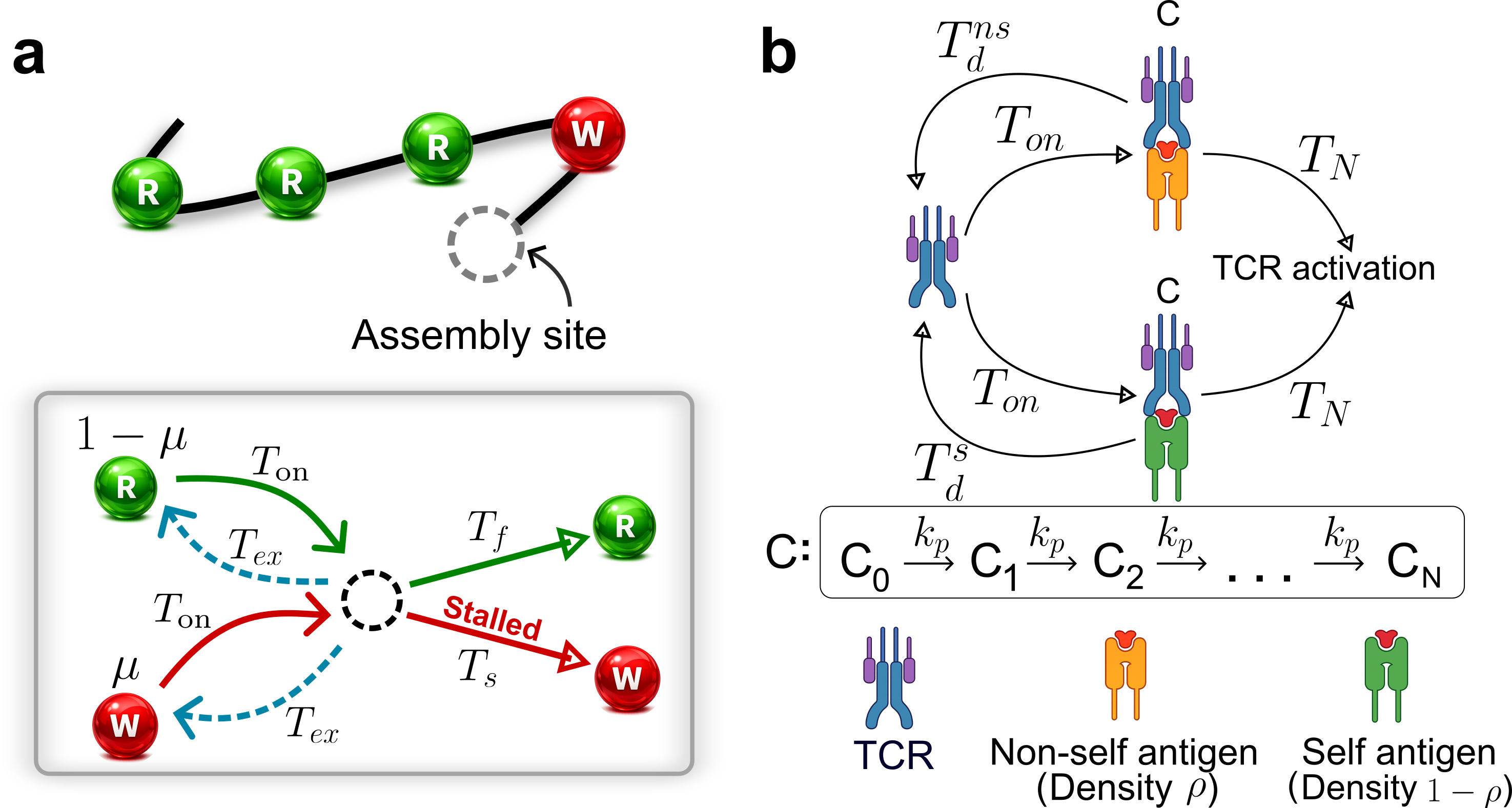}
    \caption{\textbf{a} DNA replication as an example of a self-assembly process.  A mismatched subunit `W' in the self-assembly of polymers, such as a mutation in DNA replication, leads to the stalling of the process so that subsequent addition of the subunit is hindered. Proofreading in here prevents stalling by excising the nucleotide after random time intervals $T_{ex}$, thereby facilitating the speed of the assembly process as well as reducing the error. \textbf{b} Kinetic proofreading for TCR activation network. An accurate response is generated when the TCR is activated by the non-self antigen. In here, intermittent dissociation of the TCR-pMHC complex $C$ ensures that the self antigens are rejected more often.  }
    \label{fig1}
\end{figure}

To make these ideas concrete, we now formulate a minimal model of proofreading in a stalled stochastic process. The examples motivating our study include replication, transcription, self-assembly, and immune recognition, all of which share a common architecture, as shown in \fref{fig1}. In each case, the system progresses through a sequence of states toward a successful outcome, while incorrect or unproductive trajectories generate delays and may be removed by a proofreading mechanism that effectively resets the process. In replication and self-assembly, this corresponds to the excision of an incorrectly incorporated unit; in T-cell activation, it corresponds to the dissociation of receptor–ligand complexes before successful signaling. Despite their mechanistic differences, both situations can be described within the same renewal framework, as we now proceed to show.

\emph{\textbf{Fast proofreading in replication.}---}  We first consider a minimal model of replication or self-assembly \cite{bennett1982thermodynamics}. If $\mu$ is the intrinsic error rate of the assembly process, then the assembly site attaches to it a correct subunit `R' with a probability $1-\mu$ and a wrong subunit `W' with probability $\mu$.  We assume there is a random overhead time $T_{\text{on}}$ that is taken by each subunit to get to the assembly site, which depends on the concentrations of free subunits in the surrounding environment. If the incoming subunit is the right one, it consumes a forward waiting time $T_f$ before the next subunit can be attached. If the incorporated unit is a mismatch, the configuration gets stalled, causing a significantly high forward waiting time, denoted by the stall time $T_s \gg T_f$. Evidently, for very high error rates, the replication process would be very slow due to the accumulation of significantly high stall times. 

Now consider an alternate scenario where the last incorporated subunit is excised from the system by the proofreading activity (the dashed arrows in \fref{fig1}\textbf{a}). For DNA replication, proofreading is performed by the exonuclease activity of the polymerase \cite{kunkel2004dna}, whereas for self-assembly, a slight unbinding rate of the subunit helps to avoid kinetic traps \cite{hagan2011mechanisms}. As a result of such a removal, the assembly site becomes empty again, causing the entire process to be reset. We assume that the probability of excision is the same for both the correct and wrong subunit and happens after random time intervals $T_{ex}$ \footnote{In \cite{ravasio2026conditioning}, the authors discuss the particular case of excision after periodic intervals $T_r$, which corresponds to the limit  $ f_{T_{ex}}(t)=\delta(t-T_r)$ in our approach.}. However, we note that the addition/removal of a wrong unit is more likely to be affected by the proofreading because of the associated large stall time, relative to the addition/removal of the correct unit. This immediately allows us to see that the trade-off between the error correction time and the stall time statistics decides the error rate and replication time of the genome, which we now quantify.

\textit{Error}: In addition to the basal error rate of the assembly process $\mu$, we denote by $\mu_{ec}$  the probability of a wrong subunit being incorporated even post error-correction. Then the exact error rate resulting from the proofreading follows from a renewal equation via the expression (see \fref{fig1}\textbf{a})
\begin{align}
    &\mu_{ec}=\mu \text{Pr} (T_s<T_{ex}) +\nonumber\\
    & \hspace{0.6cm}\left[\mu \text{Pr}(T_{ex}<T_s)+ (1-\mu)\text{Pr}(T_{ex}<T_f) \right]\mu_{ec}, \label{ren-err}
\end{align}
where the probability of the random variable $U$ being less than $V$ can be evaluated from their respective distributions as $\Pr (U<V)=\int_0^\infty dt f_U(t) \int_t^\infty dt' f_V(t')$ (Sec. S2 in \cite{SI}).  The first term in (1) corresponds to  events when the proofreading does not excise the incorporated wrong unit so that the error persists,  and occurs with probability  $\text{Pr} (T_s<T_{ex})$. The second composite term is the contribution when proofreading reduces errors: the first term in the bracket is the probability that a wrong subunit is incorporated and subsequently removed, while the second term takes into account the case when a correct subunit is attached and then removed. Both these terms are weighted by the possibility of getting another wrong subunit $\mu_{ec}$, after the excision of the last added subunit. 
After a slight rearrangement of \eref{ren-err}, we get the exact expression for the error rate post error correction as 
\begin{align}
    \mu_{ec}=\frac{\mu \text{Pr} (T_s<T_{ex})}{1-\mu \text{Pr}(T_{ex}<T_s)-(1-\mu)\text{Pr}(T_{ex}<T_f)}. \label{err}
\end{align}
This allows us to quantify the mean elongation time for one subunit, which we now turn to.

\textit{Mean first passage time (MFPT)}: Denoting by $T_{ec}$ the random time taken to add one subunit after error correction, we can write a renewal equation for it (see Fig. S1 in \cite{SI}) using ideas from first passage resetting \cite{evans_diffusion_2011,kusmierz2014first,reuveni_optimal_2016,pal_first_2017,pal_search_2020,reuveni2014role} as
\begin{equation}
T_{ec} = T_{\text{on}}+ \left\{
\begin{aligned}
    &\left. \begin{array}{ll} 
        T_{s} & \text{if } T_s < T_{ex} \\ 
        T_{ex} + T_{ec}' & \text{if } T_{ex} \leq T_s 
    \end{array} \right\}~ \text{Pr}(\mu), \\
    &\left. \begin{array}{ll} 
        T_f & \text{if } T_f < T_{ex} \\ 
        T_{ex} + T_{ec}'' & \text{if } T_{ex} \leq T_f 
    \end{array} \right\}~ \text{Pr}(1-\mu)
\end{aligned},
\right.
\label{renewal}
\end{equation}
where $\{T_{ec}',T_{ec}''\}$ are independent and identical (i.i.d.) copies of the random variables  $T_{ec}$.
In the above expression, the term $T_{\text{on}}$ is the minimum threshold time taken by any subunit to attach to the polymer end. The first two conditions in \eref{renewal} follow from the fact that attached subunit has a probability $\mu$ of being a mismatch; the very first condition corresponds to the case when the wrong unit remains in the system despite proofreading, which happens when $T_s<T_{ex}$, and causes an additional stall time $T_s$. Alternatively, when the stall time is higher ($T_s\ge T_{ex}$), the unit is excised, and the site becomes empty. This process consumes at least an amount of time $T_{ex}$ after which the assembly process starts afresh, which takes time $T_{ec}'$, which is an i.i.d. copy of $T_{ec}$. In a similar vein, the last two conditions can be explained by recognizing that the correct subunit is added with probability $1-\mu$ and the stall time is replaced by the forward waiting time $T_f$. To find the mean first passage time (MFPT) to add one subunit (which is inversely proportional to the average speed of the replication process), we can then calculate the expectation of \eref{renewal} and arrive at (Sec. S1 in \cite{SI})
\begin{align}
    \langle T_{ec} \rangle=\frac{\langle T_{\text{on}} \rangle+\mu \langle \text{min} (T_s, T_{ex}) \rangle + (1-\mu)\langle \text{min} (T_f, T_{ex}) \rangle }{1-\mu \text{Pr}(T_{ex}<T_s)-(1-\mu)\text{Pr}(T_{ex}<T_f)}, \label{mfpt}
\end{align}
where $\text{min}(U,V)$ denotes the minimum of the two arbitrary random variables $U$ and $V$, the mean of which can be found as $\langle \text{min}(U,V) \rangle=\int_0^\infty dt \left(\int_t^\infty dt_1 f_U(t_1)\right)\left(\int_t^\infty dt_2 f_V(t_2)\right)$  (Sec. S2 in \cite{SI}). 

We note that the expressions for the error rate given by \eref{err} and MFPT given by \eref{mfpt} hold for any generic choice of the dwell time distributions of the subunits (i.e. $f_{T_s}(t)$ and $f_{T_f}(t)$).  This allows us to quantify speed and accuracy of assembly as a function of the distribution of stall times. This simply and immediately generalizes results that are limited to the exponential distribution (where the mean and fluctuation are the same),  with constant rate $k_{stall}$ (see \cite{SI} Sec. 3).  As we show below, fluctuation in the stall time plays a crucial role in determining both speed and accuracy.


\emph{{When `more accurate is faster'?}---} 
To determine the conditions under which both the speed and accuracy could be improved, we look at the assembly process without error correction and then introduce a very small proofreading activity to the system to check how the speed and accuracy are altered \cite{reuveni_optimal_2016,pal_first_2017}.  For ease of computation, we start with by considering exponential distributions of the excision time (with rate $k_{ex}$) for the following discussion, and then allow $k_{ex}\to 0$ and show that there is no effect of excision rate on our results which holds for arbitrary stall time distributions. 

Incorporating the exponential distribution of excision times, i.e. $f_{T_{ex}}(t)=k_{ex} e^{-k_{ex} t}$ in \eref{err}, one arrives at (see Sec. S2 in \cite {SI} for details)
\begin{align}
     \mu_{ec}=\frac{\widetilde{T}_{s}(k_{ex})}{\widetilde{T}_{s}(k_{ex})+p_{\mu}\widetilde{T}_{f}(k_{ex})}, \label{err-exp}
\end{align}
where $p_{\mu}=\frac{1-\mu}{\mu}$ and $\widetilde{T}_{Z}(s)=\int_0^\infty dt f_Z(t)e^{-st}dt$ is the Laplace transform of the distribution of a generic random variable $Z$. Expansion of $\mu_{ec}$ in the limit $k_{ex}\to 0$ gives (see S4 in \cite {SI})
\begin{align}
    \mu_{ec} \approx \mu + \mu (1-\mu) \left[\langle T_f \rangle-\langle T_s \rangle\right]k_{ex} + \mathcal{O}(k_{ex}^2).
\end{align}
For the error rate to reduce with proofreading, the contribution from the correction term above must be negative, i.e.
\begin{align}
    \langle T_s \rangle >\langle T_f \rangle. \label{err-cond}
\end{align}
This intuitive result says that when the stall time is larger than the forward waiting time, proofreading will always help to reduce the error rate, irrespective of the intrinsic error rate of the process. We now turn to the condition for the MFPT of the process to reduce. Under the asumption of the exponential excision time with rate $k_{ex}$, the  MFPT given by \eref{mfpt} takes the form 
\begin{align}
    \langle T_{ec} \rangle=\frac{(1+p_\mu)(1+k_{ex}\langle T_{on} \rangle)-\widetilde{T}_{s}(k_{ex})-p_{\mu}\widetilde{T}_{f}(k_{ex})}{k_{ex}[\widetilde{T}_{s}(k_{ex})+p_{\mu}\widetilde{T}_{f}(k_{ex})]}.\label{mfpt-exp}
\end{align}
As before, we take the limit of $k_{ex} \to 0$ to obtain  (see S4  \cite{SI} )
\begin{align}
    \langle T_{ec} \rangle \approx \langle T \rangle + \left.\frac{\partial  \langle T_{ec} \rangle}{\partial k_{ex}}\right|_{k_{ex}=0} k_{ex} + \mathcal{O}(k_{ex}^2),
\end{align}
where $ \langle T \rangle=\mu \langle T_s \rangle +(1-\mu)\langle T_f \rangle$ is the MFPT without error-correction. For MFPT to reduce with proofreading we must have $\left.\frac{\partial  \langle T_{ec} \rangle}{\partial k_{ex}}\right|_{k_{ex}=0}<0$ which leads to the condition (see Sec. S5 \cite{SI})
\begin{align}
    &CV_s^2 +a^2p_\mu CV_f^2 >2\frac{\langle T_{on}\rangle}{\langle T_s\rangle}(1+a p_\mu) +\frac{2}{1+p_\mu}(1+ap_\mu)^2 \nonumber\\
    &\hspace{6cm}-a^2p_\mu-1, \label{mfpt-cond}
\end{align}
where $a=\frac{\langle T_f \rangle}{\langle T_s \rangle}$ and the coefficient of variation $CV_Z=\frac{\sqrt{\langle Z^2 \rangle-\langle Z \rangle^2}}{\langle Z \rangle }$ is the usual ratio of fluctuation to mean of the random variable $Z$.  

Together, \eref{err-cond} and \eref{mfpt-cond} are our central results which decide the \textit{sufficient} conditions (see End Matter) to determine when the process of proofreading (by rejecting potentially stalled processes) can improve speed and accuracy. The first condition \eref{err-cond} decides when the error rate can be improved, and the second  \eref{mfpt-cond} decides whether MFPT could be reduced as well. In the limit when the mean stall time is sufficiently large compared to the other system timescales in the problem, i.e. when $\langle T_s \rangle \gg \langle T_f \rangle, \langle T_{on} \rangle$, corresponding to $a p_\mu \to 0$, the expression \eref{mfpt-cond} simplifies to
\begin{align}
    CV_s > \sqrt{2\mu-1}. \label{cond}
\end{align}
This inequality highlights the interplay between the intrinsic error rate and the fluctuations of stalled states. In the strongly stalled regime, proofreading can simultaneously improve speed and accuracy whenever the variability of the stall time exceeds the threshold set by Eq.~(\ref{cond}). For $\mu<1/2$, this condition is automatically satisfied, whereas for larger error rates sufficiently broad stall-time distributions are required. Exponentially distributed stall times correspond to $CV_s=1$ \cite{ravasio2026evolution}, while the bi-exponential distributions observed in several experiments \cite{hoekstra2017switching,neuman2003ubiquitous,shaevitz2003backtracking,yuan2025mismatched} satisfy $CV_s>1$, making both classes naturally compatible with the ``more accurate is faster'' regime.

\begin{figure}
    \centering
    \includegraphics[width=8.6cm]{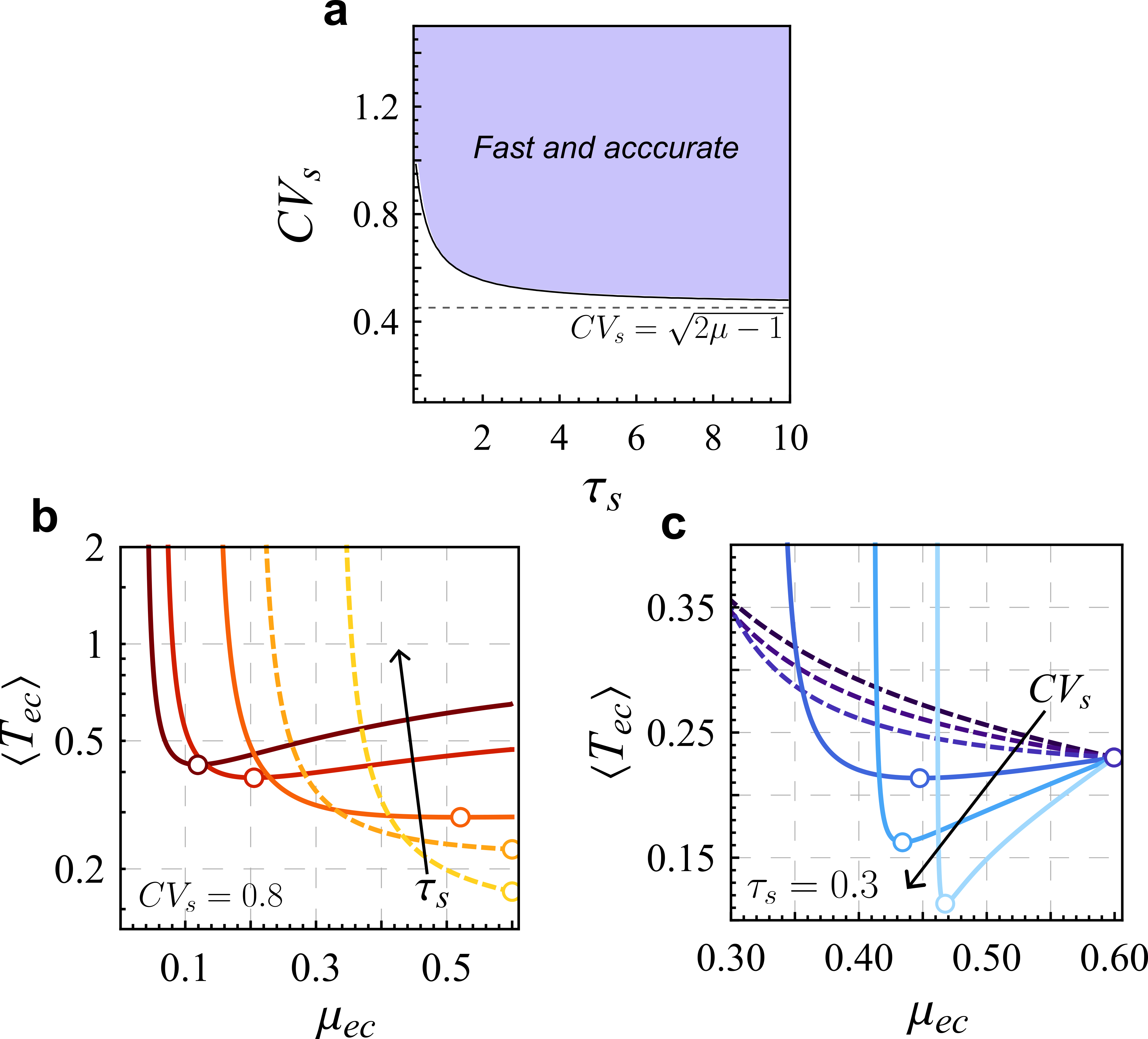}
    \caption{\textbf{Illustration for gamma distributed stall times, when proofreading improves speed and accuracy simultaneously.} \textbf{a} Parameter space (shaded region) where the conditions \eref{err-cond} and \eref{mfpt-cond} holds (for $\mu=0.6$). In the limit of $\langle T_s \rangle \gg \langle T_f \rangle$ the separatrix is given by $CV_s =\sqrt{2\mu-1}$. \textbf{b} We fix $CV_s=0.8$ and vary $\tau_s \in [0.2,1]$. We see that as $\tau_s$ increases beyond $\tau_s=0.38$ as given by the separatrix in \textbf{a}, we get a region where both the error ($\mu_{ec}$) and MFPT ($\langle T_{ec} \rangle$) decrease (region right to the circles in each curve). The curves where the criteria (\eref{mfpt-cond}) are not satisfied is marked by the dashed lines.  We choose a dashed curve from \textbf{b} with fixed $\tau_s=0.3$ and vary $CV_s \in[0.6,1.4]$ in \textbf{c}. We see that when $CV>0.87$, we again recover the faster and more accurate regime.
}
    \label{fig2}
\end{figure}

Having identified the coefficient of variation as a key control parameter, we now use Gamma-distributed waiting times to isolate its effect, as the mean and coefficient of variation can be tuned independently in this distribution. We consider stall times distributed as
\begin{align}
    T \sim f(\alpha,\beta,t)=\frac{\beta^\alpha}{\Gamma(\alpha)} t^{\alpha-1}e^{-\beta t},
\end{align}
so that $\langle T \rangle=\frac{\alpha}{\beta}$ and $CV_T=\frac{1}{\sqrt{\alpha}}$. To illustrate the role of fluctuations, we consider forward and stall times parameterized in terms of the mean stall time $\tau_s=\langle T_s\rangle$ and coefficient of variation $CV_s$ which can be varied independently while keeping $\tau_f=\langle T_f\rangle=0.1$ fixed. Figure~\ref{fig2}a shows the region where Eqs.~(\ref{err-cond}) and (\ref{mfpt-cond}) are simultaneously satisfied, corresponding to the ``fast and accurate'' regime. Two distinct routes lead into this regime. First, increasing the mean stall time beyond a threshold ($\tau_s>0.38$ for $CV_s=0.8$) causes the MFPT--error curves to become non-monotonic, creating a region where proofreading simultaneously decreases both error and completion time (Fig.~\ref{fig2}b). Second, even when the mean stall time is fixed ($\tau_s=0.3$), increasing the fluctuation level beyond $CV_s>0.87$ produces the same transition (Fig.~\ref{fig2}b). Thus, both the average cost of stalled trajectories and their variability govern the emergence of the ``more accurate is faster’’ regime. It may be pertinent to note that Gamma-distributed waiting times arise naturally in multi-step reaction networks and have been observed in several biological systems, including replication, translation, and immune signaling \cite{bera2021nucleotide,wen2008following,dulin2015elongation,huang2019molecular,johnson2017single}, so that our results may have direct implications there.

\emph{\textbf{Fast proofreading of T-cell response}---} We now turn to briefly discuss when T-cell response in the immune system can be fast and accurate, following the adaptation of kinetic proofreading for this process  \cite{mckeithan1995kinetic,goldstein2004mathematical}. It is known that T-cells (T lymphocytes) play a governing role in the adaptive immune response by detecting a few non-self antigens from a pool of self antigens in the antigen-presenting cells (APC). They do this using T-cell receptors (TCR) that bind to the peptides presented by the major histocompatibility complex (pMHC) presented by a self or non-self antigen in the APC. To detect an infection efficiently, the TCRs must activate or send a signal upon detection of non-self antigens in the APC that must be very fast (to detect the pathogen) and highly accurate (reducing the risk of autoimmune infections).  The classical McKeithan model \cite{mckeithan1995kinetic} suggests that the TCR-pMHC bond dwell time $T_d$ plays a pivotal role in deciding the discriminatory power of TCRs, whereby each TCR-pMHC complex ($C$) is formed after random time intervals $T_{\text{on}}$ (\fref{fig1}\textbf{b}). Upon bond formation, a sequence of intermediate cascaded reactions (each with a forward rate $k_p$) is initiated. Each of these reactions produces a new intermediate complex $C_i$, with $i \in [0,N]$. For any signalling to happen from a particular TCR-pMHC bond, the random dwell time $T_d$ must be greater than the time required for proofreading steps to reach the final step $N$, which we denote by $T_N$. 

The accuracy of T-cell response is generally measured by two metrics: specificity, i.e. the probability that no signalling occurs when bound with a self-antigen, and sensitivity, i.e. the probability of signalling when bound with a non-self one. If $T_d^{s}$ and $T_d^{ns}$, denotes the random dwell time of self and non-self antigens on the TCR, respectively, then  specificity is correlated with $\text{Pr}(T_d^{s}<T_N)$ and sensitivity is correlated with $\text{Pr}(T_N<T_d^{ns})$. Then, overall accuracy is quantified as $\rho \text{Pr}(T_N<T_d^{ns}) +(1-\rho)\text{Pr}(T_d^{s}<T_N)$, where $\rho$ and $1-\rho$ are the density of non-self and self antigens, respectively. This allows us to write the error rate  as
\begin{align}
    \eta=1-\rho \text{Pr}(T_N<T_d^{ns}) -(1-\rho)\text{Pr}(T_d^{s}<T_N). \label{tcell-err}
\end{align}
We can write a renewal equation for the random time of activation of a TCR, i.e. $ T_{act}$, in a similar fashion to \eref{renewal}, to arrive at the result for the mean activation time (see SI for a detailed derivation)
\begin{align}
     \langle T_{act}\rangle= \frac{\langle T_{on} \rangle + \rho \langle \text{min}(T_N, T_d^{ns}) \rangle + (1-\rho) \langle \text{min}(T_N, T_d^{s}) \rangle}{1-\rho \Pr (T_d^{ns}\leq T_N) -(1-\rho)\Pr (T_d^{s}\leq T_N)}. \label{tcell-mfpt}
\end{align}

\begin{figure}[b]
    \centering
    \includegraphics[width=5cm]{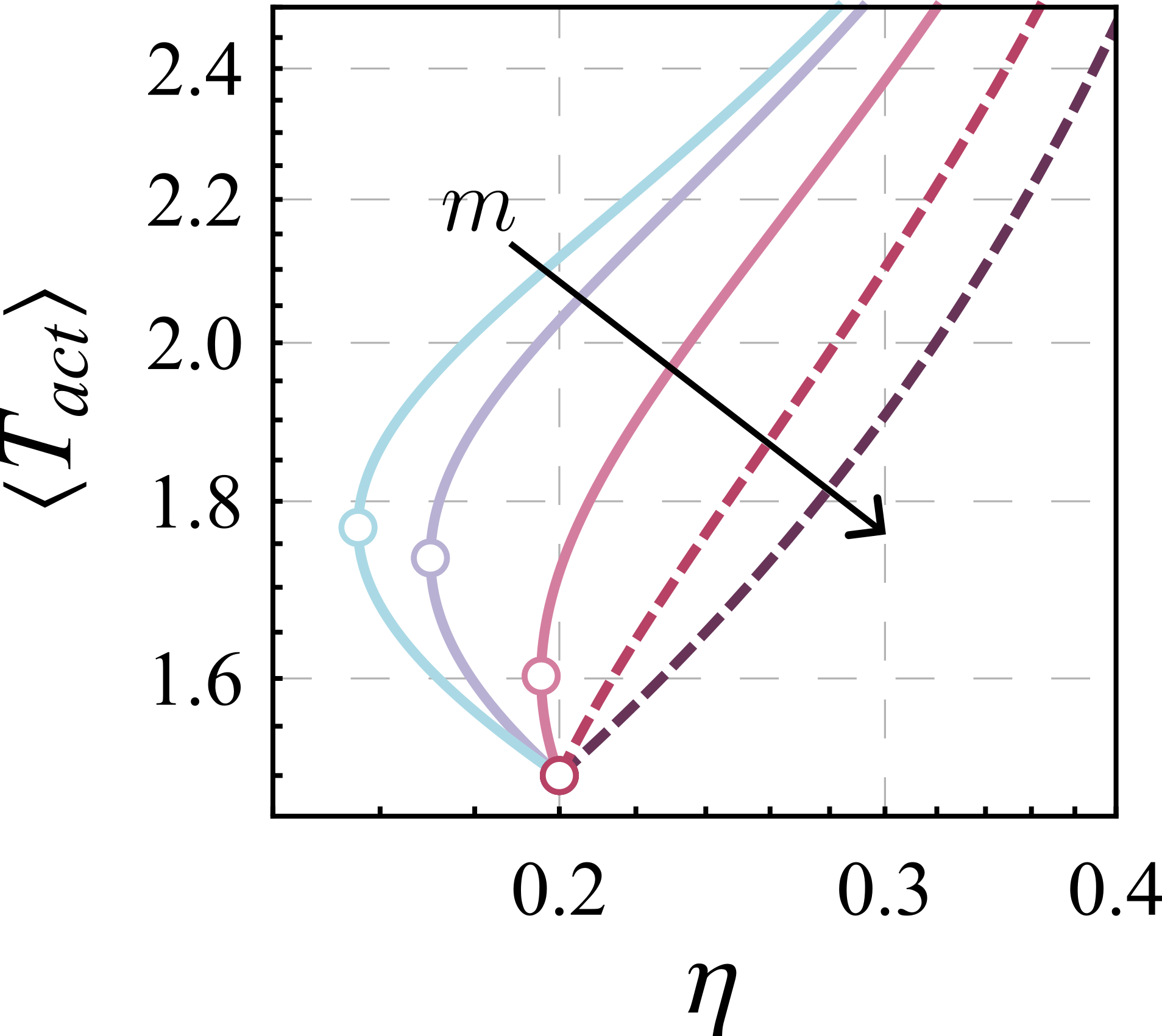}
    \caption{\textbf{`Fast and accurate' T-cell activation -} The region above the circles in the solid curves is where `more accurate is faster' for T-cell activation governed by \eref{cond-1} and \eref{cond-2}. The curves undergo a behavioural transition from a non-monotonic to a monotonic variation with increasing $m=\frac{\langle T_d^{s}\rangle}{\langle T_d^{ns}\rangle}$. Other parameters are: $N=5, k_p=10, \rho=0.8$.}
    \label{fig33}
\end{figure}

As for the case of replication, we assume that the dwell time distribution for self and non-self antigens comes from an exponential distribution with rate $k_{\text{off}}$ and $m k_{\text{off}}$, respectively, where $m<1$, and then take consider the limit $k_{\rm off} \to 0$ to show that this assumption does not change our qualitative results. Substituting this dwell time distribution in \eref{tcell-err} and \eref{tcell-mfpt} and taking the limit $k_{\text{off}}\to 0$ we arrive at the following conditions for the MFPT and error to simultaneously decrease (see Sec. S7 in \cite{SI}):
\begin{align}
    &CV_{T_N}^2>1+2\frac{\langle T_{on}\rangle}{\langle T_N\rangle}, \label{cond-1}\\
  \text{and,}~~~  & m<\frac{1-\rho}{\rho}<1.\label{cond-2}
\end{align}

As a concrete example, for the immune reaction network as shown in \fref{fig1}\textbf{b}, $T_N$ comes from an Erlang distribution which is a gamma distribution with integer values of shape parameters i.e. $f(N,k_p,t)$ with $CV_{T_N}=\frac{1}{\sqrt{N}}$, so that \eref{cond-1} is never satisfied. This implies that increasing $k_{\text{off}}$ will only increase the mean activation time. However, for a suitable choice of $m$ via \eref{cond-2}, the accuracy can either increase or decrease with $k_{\text{off}}$, resulting in a non-monotonic variation in the MFPT-error plot as shown in \fref{fig33} for $\rho=0.8$. The region above the circle for each curve is where both accuracy and speed are positively correlated. As we increase $m$ in the range $m\in [\frac{1}{10},\frac{1}{2}]$, we transition to a regime where a decrease in error comes with an increase in the MFPT, with the exact transition point being $m=\frac{1}{4}$ as determined from \eref{cond-2}.

\emph{\textbf{Discussion}---}  Using a non-Markovian renewal approach, we have derived general conditions under which proofreading simultaneously improves speed and accuracy. In a minimal setting, this allows us to show that the key control parameters are the intrinsic error rate together with the mean and fluctuations of the stall time.  Our results show that the benefit of proofreading is determined not only by the average duration of stalled states, but also by their variability. Long and highly variable stalls make unsuccessful trajectories disproportionately costly in  replication, self-assembly, and immune recognition, so that resetting can both reduce errors and accelerate completion. The resulting criteria provide a simple way to assess when proofreading is advantageous without requiring detailed knowledge of the underlying microscopic mechanisms. Beyond these specific examples, it suggests a broader perspective in which kinetic proofreading may be viewed as a biological realization of stochastic resetting. Understanding how the statistics of stalled states are regulated could therefore provide new insight into the design principles of nonequilibrium information-processing systems.




\emph{\textbf{Acknowledgments.}---}  AB thanks USIEF and IIE for the Fulbright fellowship at Harvard during which the project was completed. LM thanks the Simons Foundation and the Henri Seydoux Fund for partial financial support.  We thank A. Murugan for alerting us to the study \cite{ravasio2026conditioning} just as our work was being completed; while their work discusses the connection to resetting, it does not take the renewal approach outlined here.\\

\emph{\textbf{Data availability}---} The data that support the findings of this article are openly available \cite{key}.

\let\savedaddcontentsline\addcontentsline
\renewcommand{\addcontentsline}[3]{}
\let\savedaddtocontents\addtocontents
\renewcommand{\addtocontents}[2]{}

\bibliography{fpusr}

\newpage

\section*{End Matter}
\begin{figure}[t]
    \centering
    \includegraphics[width=9.6cm]{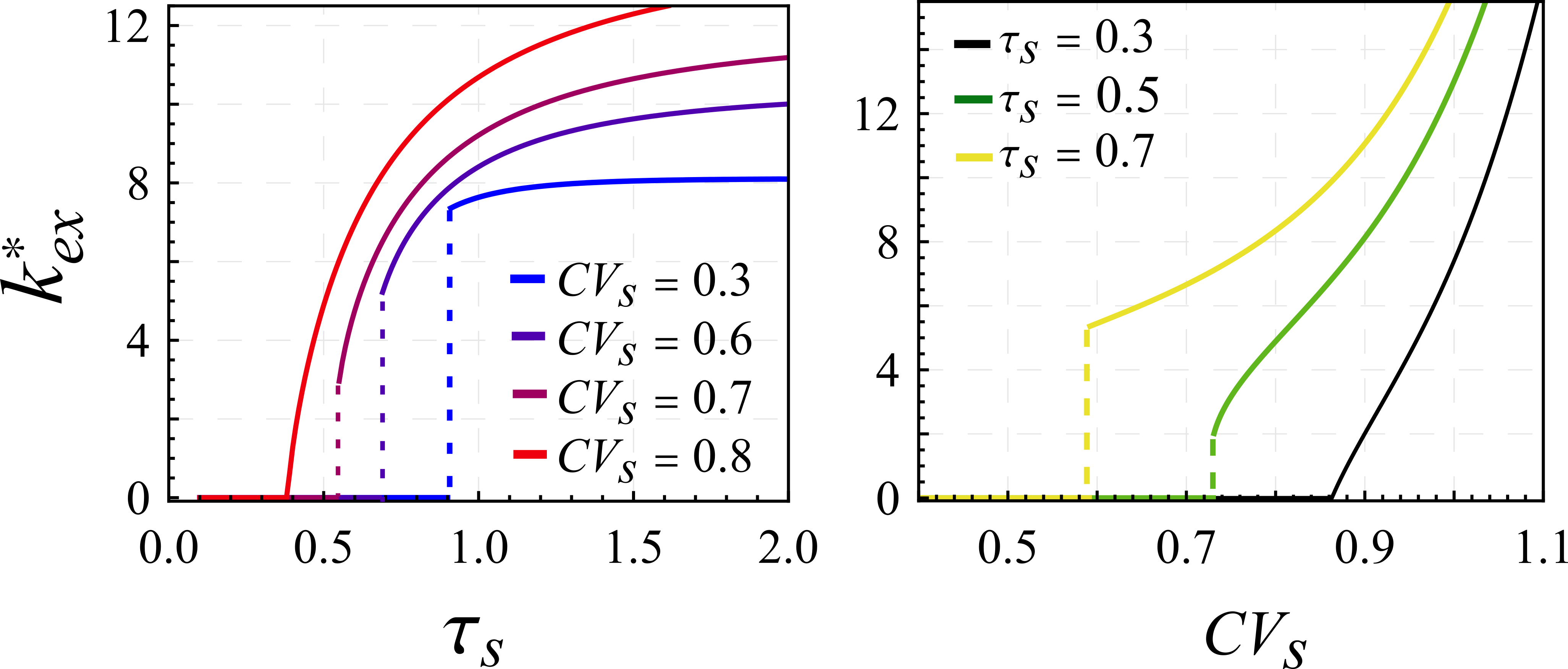}
    \caption{\textbf{Nature of the transition}: The optimal value of rate of excision as obtained from \eref{orr} exhibits both first and second order phase transitions with respect to $\tau_s$ and $CV_s$. For first-order transition, however, the criterion in \eref{mfpt-cond} does not necessarily hold.}
    \label{fig3}
\end{figure}

\emph{\textbf{Sufficient but not necessary conditions}---} We wish to now establish that the criteria in \eref{err-cond} and \eref{mfpt-cond}, although sufficient in determining the `accurate and fast proofreading' regime, are not necessary ones. To see this, we first quantify the value of the optimal excision rate  $k_{ex}^*$ at which the MFPT is lowest as observed in \fref{fig2}(\textbf{b,c}) (shown by the circles), found using the condition
\begin{align}
    \left.\frac{\partial\langle T_{ec}\rangle}{\partial k_{ex}}\right|_{k_{ex}=k_{ex}^*}=0. \label{orr}
\end{align}
Any non-zero value of $k_{ex}^*$ implies that proofreading can help to increase the speed of the assembly process \cite{pal2019landau}. For example, with the value of $CV_s=0.8$ as in \fref{fig2}\textbf{b}, we see that $k_{ex}^*$
exhibits a transition (corresponding to a supercritical bifurcation or a second-order type) whereby the curves become non-monotonic as $\tau_s$ is varied, as shown by the solid red line in \fref{fig3}\textbf{a}. A similar transition is observed when we vary $CV_s$ keeping $\tau_s$ fixed at $0.3$ as observed in \fref{fig2}\textbf{c} earlier, and the transition of $k_{ex}^*$ is shown by the solid black curve in \fref{fig3}\textbf{b}. For such second-order transitions, the $CV$ criteria as in \eref{mfpt-cond} is sufficient  to evaluate the exact regimes when one can obtain faster and more accurate results. 

However, there are regimes where $k_{ex}^*$ can also undergo a first-order (subcritical bifurcation) transition both with respect to $\tau_s$ and $CV_s$, as shown in \fref{fig3}(\textbf{a,b}). Since the $k_{ex}^*$ jumps to a large finite value at the transition points, the limit $k_{ex} \to 0$ that leads to \eref{mfpt-cond} does not give a correct result. Thus, \eref{err-cond} and \eref{mfpt-cond} constitute a sufficient but not necessary condition for fast and accurate proofreading.

\newpage

\let\oldlabel\label      
\renewcommand{\label}[1]{} 
\begin{titlepage}
\title{Supplemental material for \\ \underline{``When proofreading improves speed and accuracy"}}
\maketitle
\end{titlepage}
\let\label\oldlabel


\onecolumngrid
\setcounter{page}{1}
\renewcommand{\thepage}{S\arabic{page}}
\setcounter{equation}{0}
\renewcommand{\theequation}{S\arabic{equation}}
\setcounter{figure}{0}
\renewcommand{\thefigure}{S\arabic{figure}}
\setcounter{section}{0}
\renewcommand{\thesection}{S\arabic{section}}
\setcounter{table}{0}
\renewcommand{\thetable}{S\arabic{table}}

This Supplemental Material provides detailed derivations of some of our main results and provides additional discussions that support our findings.

\let\addcontentsline\savedaddcontentsline
\let\addtocontents\savedaddtocontents

\tableofcontents
\section{Schematic depiction of the renewal process}

\begin{figure}[H]
    \centering
    \includegraphics[width=16cm]{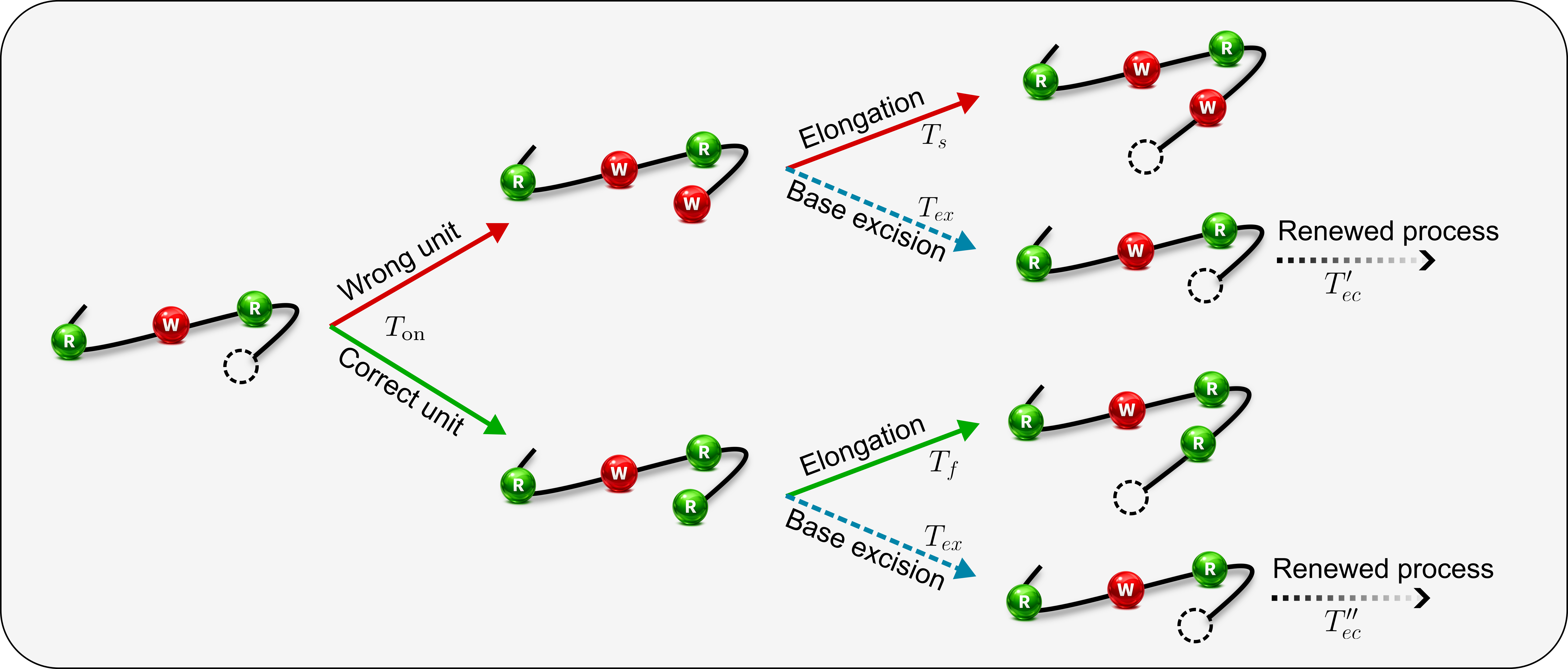}
    \caption{In here, we show 4 different outcomes of the self-assembly process for DNA replication - i) addition of a correct nucleotide `R' and the polymerase translocates to the next elongation site, ii) addition of a correct nucleotide `R', and the exonuclease activity removes the unit, essentially renewing the process, iii) addition of a wrong nucleotide `W' and the polymerase translocates to the next elongation site, iv) addition of a wrong nucleotide `W' and the exonuclease activity removes the unit which again renews the process. All these possibilities are accounted for in the renewal equation (3) of the main text.}
    \label{figs1}
\end{figure}
We recall that, the renewal equation for the above process is written as 
\begin{equation}
T_{ec} = T_{\text{on}}+ \left\{
\begin{aligned}
    &\left. \begin{array}{ll} 
        T_{s} & \text{if } T_s < T_{ex} \\ 
        T_{ex} + T_{ec}' & \text{if } T_{ex} \leq T_s 
    \end{array} \right\}~ \text{Pr}(\mu), \\
    &\left. \begin{array}{ll} 
        T_f & \text{if } T_f < T_{ex} \\ 
        T_{ex} + T_{ec}'' & \text{if } T_{ex} \leq T_f 
    \end{array} \right\}~ \text{Pr}(1-\mu)
\end{aligned}.
\right.
\end{equation}
Each of the branches of the \fref{figs1} contributes to each line of the renewal equation. The above renewal equation can also be written as
\begin{align}
    T_{ec} &=T_{on} + \Pr(\mu)\times \left[ I( T_s < T_{ex}) T_s +  I( T_{ex} \le T_s) (T_{ex}+T_{ec}')\right] \nonumber\\
    &\hspace{7cm}+ \Pr(1-\mu)\times \left[ I( T_f < T_{ex}) T_f +  I( T_{ex} \le T_f) (T_{ex}+T_{ec}'')\right], \nonumber\\
    &=T_{on} + \Pr(\mu)\times \left[ \text{min} (T_s, T_{ex}) + I( T_{ex} \le T_s)T_{ec}'\right] + \Pr(1-\mu)\times \left[ \text{min} (T_f, T_{ex}) + I( T_{ex} \le T_f)T_{ec}''\right], \nonumber\\
\end{align}
where $I(\text{condition})$ is an indicator function which takes the value unity only when the condition is true, zero otherwise. We also denote by $\text{min}(U,V)$ as the minimum of two random variables $U$ and $V$ found as $\text{min}(U,V)=I(U<V) U+ I(V \le U) V$. After taking an expectation of the renewal equation over all the different distributions of different random variables, denoted by $\langle ... \rangle$, we have
\begin{align}
    \langle T_{ec} \rangle &=\langle T_{on} \rangle + \mu\times \left[ \langle \text{min} (T_s, T_{ex}) \rangle + \langle I( T_{ex} \le T_s)\rangle \langle T_{ec}' \rangle\right] + (1-\mu)\times \left[ \langle \text{min} (T_f, T_{ex}) \rangle + \langle I( T_{ex} \le T_f)\rangle \langle T_{ec}'' \rangle\right].
\end{align}
We now note that the expectation of the indicator function is just the probability such $\langle I (U<V) \rangle=\Pr(U<V)$. Furthermore, since  $\{T_{ec}',T_{ec}''\}$ are independent and identical copies of $T_{ec}$ we have $ \langle T_{ec}' \rangle= \langle T_{ec}'' \rangle= \langle T_{ec} \rangle$. The above two observations finally lead to the result after a slight rearrangement
\begin{align}
    \langle T_{ec} \rangle=\frac{\langle T_{\text{on}} \rangle+\mu \langle \text{min} (T_s, T_{ex}) \rangle + (1-\mu)\langle \text{min} (T_f, T_{ex}) \rangle }{1-\mu \text{Pr}(T_{ex}<T_s)-(1-\mu)\text{Pr}(T_{ex}<T_f)}, 
\end{align}
as shown in Eq. (3) of the main text.

\section{Derivation of  $\text{Pr}(U<V)$ and  $\langle \text{min} (U, V) \rangle$}
The general results for the error rate (\eref{err}) MFPT (\eref{mfpt}) in main text requires evaluation of quantities such as $\langle \text{min} (U, V) \rangle$ and $\text{Pr}(U<V)$ for two aribitrary random variables $U$ and $V$. To evaluate them consider they have distributions 
\begin{align}
    U \sim f_U(t),~~V\sim f_V(t)
\end{align}
Let us first calculate the quantity  $\text{Pr}(U<V)$ as follows:
\begin{align}
     \text{Pr}(U<V)=\int_0^\infty dt f_U(t) \text{Pr}(V>t)=\int_0^\infty dt f_U(t) \left(\int_t^\infty dt' f_V(t')\right).
\end{align}
Let us now calculate the quantity $\langle \text{min} (U, V) \rangle$. Let $Z$  represent the random variable 
\begin{align}
    Z=\text{min}(U,V),
\end{align}
with the associated density $f_Z(t)$. We thus have
\begin{align}
    \langle \text{min}(U,V)\rangle=\langle Z\rangle=\int_0^\infty t f_Z(t) dt. \label{c2}
\end{align}
The density $f_Z(t)$ can be found from the cumulative distribution of $Z$ as
\begin{align}
    f_Z(t)=-\frac{d}{dt}\text{Pr}(Z>t). \label{fz}
\end{align}
Now the cumulative distribution $\text{Pr}(Z>t)$ can easily be found by noting that the condition $Z=\text{min}(U,V)>t$ holds only when both of $U$ and $V$ are simultaneously greater than $t$. That in turn implies
\begin{align}
  \text{Pr}(Z>t)=\text{Pr}(U>t)\text{Pr}(V>t). \label{pzt}
\end{align}
One can now do an integration by parts of the integral in \eref{c2} to have
\begin{align}
    \langle \text{min}(U,V)\rangle&=\int_0^\infty t f_Z(t) dt\nonumber\\
    &=\left[t\int f_Z(t)dt\right]_0^\infty-\int_{0}^\infty dt \left(\int dt f_Z(t)\right).
\end{align}
Upon inserting \eref{fz} to the above equation, we have
\begin{align}
     \langle \text{min}(U,V)\rangle&=-\lim_{t\to \infty}t\text{Pr}(Z>t)+\int_0^\infty dt\text{Pr}(Z>t).
\end{align}
Assuming $\text{Pr}(Z>t)$ to decay faster than $1/t$ as $t\to\infty$, we finally have
\begin{align}
    \langle \text{min}(U,V)\rangle=\int_0^\infty dt\text{Pr}(Z>t),
\end{align}
which upon substituting result form \eref{pzt} yields
\begin{align}
     \langle \text{min}(U,V)\rangle=\int_0^\infty dt \text{Pr}(U>t)\text{Pr}(V>t). \label{mintr1}
\end{align}
Now from the distribution of $U$ and $V$ we have
\begin{align}
    &\text{Pr}(U>t)=\int_t^{\infty} f_U(t')dt', \\
    & \text{Pr}(V>t)=\int_t^\infty f_V(t')dt'
\end{align}
Plugging these results in \eref{mintr1} we have 
\begin{align}
      \langle \text{min}(U,V) \rangle &= \int_0^\infty dt\left(\int_t^\infty dt'f_U(t')\right) \left(\int_t^\infty dt'f_    V(t')\right).
\end{align}
As a special case, when $V$ comes from an exponential distribution with rate $k_{ex}$ such that $f_V(t)=k_{ex} e^{-k_{ex}t}$, we have
\begin{align}
   &  \text{Pr}(U<V)=\int_0^\infty dt f_U(t) \left(\int_t^\infty dt'k_{ex} e^{-k_{ex} t'}\right)=\int_0^\infty dt f_U(t) e^{-k_{ex} t}=\widetilde{T}_U(k_{ex}),\\
     \implies & \text{Pr}(V<U)=1-\widetilde{T}_U(k_{ex}), \label{pr-1}
\end{align}
where $\widetilde{T}_U(k_{ex})$ is the Laplce transform of the density $f_U(t)$ with the variable $k_{ex}$. Similarly, we have,
\begin{align}
      \langle \text{min}(U,V) \rangle =\int_0^\infty dt e ^{-k_{ex}t}\left(\int_t^\infty dt'f_U(t')\right) =\frac{1-\widetilde{T}_U(k_{ex})}{k_{ex}}. \label{min-1}
\end{align}
Note that \eref{pr-1} and \eref{min-1} have been used to derive the results in \eref{err-exp} and \eref{mfpt-exp}.

\section{Error and MFPT in the Markovian description}
In this section, we derive the result for the error and MFPT of the self-assembly process when all the time-scales are chosen from the exponential distribution. In particular, we chose
\begin{align}
    T_{on}\sim k_{on}e^{-k_{on}t},~~T_f \sim k_f e^{-k_f t},~~T_s \sim k_{stall}e^{-k_{stall}t},~~T_{ex}\sim k_{ex}e^{-k_{ex}t}.
\end{align}
For the above choice of parameters from Eq. (2) of the main text we have the error rate given by
\begin{align}
    \mu_{ec}&=\frac{\frac{k_{stall}}{k_{stall}+k_{ex}}}{\left(\frac{1-\mu}{\mu}\right)\frac{k_{f}}{ k_{f}+k_{ex}}+\frac{k_{stall}}{k_{stall}+k_{ex}}}.
\end{align}
To find the MFPT we employ Eq. (4) in the main text that yields
\begin{align}
    \langle T_{ec}\rangle=\frac{\mu^{-1}  \left(1+\frac{k_{ex}}{k_{on}}\right)-\frac{k_{f} (1-\mu )}{\mu  (k_{f}+r)}-\frac{k_{stall}}{k_{stall}+k_{ex}}}{k_{ex} \left(\frac{k_{f} (1-\mu )}{\mu  (k_{f}+k_{ex})}+\frac{k_{stall}}{k_{stall}+k_{ex}}\right)}.
\end{align}

\section{Derivation of condition \eref{err-cond}}
We recall the expression for the error under proofreading with a rate $k_{ex}$ as in \eref{err-exp} in the main text
\begin{align}
     \mu_{ec}=\frac{\widetilde{T}_{s}(k_{ex})}{\widetilde{T}_{s}(k_{ex})+p_{\mu}\widetilde{T}_{f}(k_{ex})}. \label{err-sm}
\end{align}
We check whether the introduction of a very small excision rate ($k_{ex}\to 0$) reduces the error. To see this, we first note
\begin{align}
    \widetilde{T}_{s}(k_{ex})=1-k_{ex}\langle T_s \rangle +\frac{k_{ex}^2}{2} \langle T_{s}^2 \rangle +\mathcal{O}(k_{ex}^2),~~\widetilde{T}_{f}(k_{ex})=1-k_{ex}\langle T_f \rangle +\frac{k_{ex}^2}{2} \langle T_{f}^2 \rangle + \mathcal{O}(k_{ex}^2). \label{expand}
\end{align}
Using the above expression in the equation \eref{err-sm} and neglecting terms $ \mathcal{O}(k_{ex}^2)$ we have
\begin{align}
     \mu_{ec}&=\frac{1-k_{ex}\langle T_s \rangle}{1-k_{ex}\langle T_s \rangle+p_{\mu}[1-k_{ex}\langle T_f \rangle]}\nonumber\\
     &=\frac{1-k_{ex}\langle T_s \rangle}{(1+p_\mu)\left[1- \frac{k_{ex}}{1+p_\mu}(\langle T_s \rangle+p_\mu \langle T_f \rangle)\right]}\nonumber\\
     &=\frac{1-k_{ex}\langle T_s \rangle}{(1+p_\mu)}\left[1+ \frac{k_{ex}}{1+p_\mu}(\langle T_s \rangle+p_\mu \langle T_f \rangle)\right]\nonumber\\
     &=\frac{1}{1+p_\mu}\left(1+ k_{ex}\left[-\langle T_s \rangle + \frac{\langle T_s \rangle+p_\mu \langle T_f \rangle}{1+p_\mu}\right]\right) \nonumber\\
     &=\frac{1}{1+p_\mu}\left(1+ \frac{k_{ex}p_\mu}{1+p_\mu}\left[-\langle T_s \rangle + \langle T_f \rangle\right]\right) \nonumber\\
     &=\mu\left(1+ \frac{k_{ex}p_\mu}{1+p_\mu}\left[-\langle T_s \rangle + \langle T_f \rangle\right]\right), \label{err-1}
\end{align}
where we identify that $\frac{1}{1+p_\mu}=\mu$. For error to reduce beyond $\mu$ the second term in \eref{err-1} must be negative which is true only when
\begin{align}
    \langle T_s \rangle>\langle T_f \rangle,
\end{align}
which is the condition as derived in \eref{err-cond} in the main text.

\section{Derivation of condition \eref{mfpt-cond}}
We start the derivation by recalling the exact result for the MFPT as in \eref{mfpt-exp} of the main text.
\begin{align}
    \langle T_{ec} \rangle=\frac{(1+p_\mu)(1+k_{ex}\langle T_{on} \rangle)-\widetilde{T}_{s}(k_{ex})-p_{\mu}\widetilde{T}_{f}(k_{ex})}{k_{ex}[\widetilde{T}_{s}(k_{ex})+p_{\mu}\widetilde{T}_{f}(k_{ex})]}. \label{mfpt-sm}
\end{align}
Likewise, the error, we ask whether the introduction of a very small excision rate ($k_{ex}\to 0$) reduces the MFPT. Plugging the results from \eref{expand} in \eref{mfpt-sm} and neglecting $\mathcal{O}(k_{ex}^2)$ we have 
\begin{align}
    \langle T_{ec} \rangle_{k_{ex}\to 0}&= \frac{(1+p_\mu)(1+k_{ex}\langle T_{on} \rangle)-1+k_{ex}\langle T_s \rangle -\frac{k_{ex}^2}{2} \langle T_{s}^2 \rangle-p_{\mu}[1-k_{ex}\langle T_f \rangle +\frac{k_{ex}^2}{2} \langle T_{f}^2 \rangle]}{k_{ex}[1-k_{ex}\langle T_s \rangle+p_{\mu}(1-k_{ex}\langle T_f \rangle)]} \nonumber\\
    &= \frac{(1+p_\mu)\langle T_{on} \rangle+\langle T_s \rangle -\frac{k_{ex}}{2} \langle T_{s}^2 \rangle-p_{\mu}[-\langle T_f \rangle +\frac{k_{ex}}{2} \langle T_{f}^2 \rangle]}{1-k_{ex}\langle T_s \rangle+p_{\mu}(1-k_{ex}\langle T_f \rangle)} \nonumber\\
    &= \frac{(1+p_\mu)\langle T_{on} \rangle+\langle T_s \rangle +p_{\mu}\langle T_f \rangle -\frac{k_{ex}}{2} [\langle T_{s}^2 \rangle+p_{\mu} \langle T_{f}^2 \rangle]}{1+p_\mu-k_{ex}[\langle T_s \rangle+p_\mu\langle T_f \rangle]} \nonumber\\
    &=\frac{(1+p_\mu)\langle T_{on} \rangle+\langle T_s \rangle +p_{\mu}\langle T_f \rangle -\frac{k_{ex}}{2} [\langle T_{s}^2 \rangle+p_{\mu} \langle T_{f}^2 \rangle]}{1+p_\mu} \times \left(1+\frac{k_{ex}}{1+p_\mu}[\langle T_s \rangle+p_\mu\langle T_f \rangle]\right) \nonumber\\
    &=\langle T_{on} \rangle+\frac{\langle T_s \rangle+p_\mu \langle T_f \rangle}{1+p_\mu} + \frac{k_{ex}}{1+p_\mu} \left(-\frac{\langle T_{s}^2 \rangle+p_{\mu} \langle T_{f}^2 \rangle}{2} + (\langle T_s \rangle+p_\mu \langle T_f \rangle)\left(\langle T_{on} \rangle+\frac{\langle T_s \rangle+p_\mu \langle T_f \rangle}{1+p_\mu}\right)\right).
\end{align}
For proofreading to reduce the MFPT, one must have the contribution from the first-order term in the above expansion be negative, such that
\begin{align}
    &-\frac{\langle T_{s}^2 \rangle+p_{\mu} \langle T_{f}^2 \rangle}{2} + (\langle T_s \rangle+p_\mu \langle T_f \rangle)\left(\langle T_{on} \rangle+\frac{\langle T_s \rangle+p_\mu \langle T_f \rangle}{1+p_\mu}\right)<0,\nonumber\\
    \implies& \langle T_{s}^2 \rangle+p_{\mu} \langle T_{f}^2 \rangle>2(\langle T_s \rangle+p_\mu \langle T_f \rangle)\left(\langle T_{on} \rangle+\frac{\langle T_s \rangle+p_\mu \langle T_f \rangle}{1+p_\mu}\right)\nonumber\\
    \implies& CV_s^2 \langle T_s \rangle^2 +p_\mu CV_f^2 \langle T_f\rangle^2>2 \langle T_{on}\rangle (\langle T_s \rangle+p_\mu \langle T_f \rangle) +\frac{2}{1+p_\mu} (\langle T_s \rangle+p_\mu \langle T_f \rangle)^2-\langle T_s \rangle^2-p_\mu\langle T_f \rangle^2\nonumber\\
\implies &CV_s^2 +a^2p_\mu CV_f^2 >2\frac{\langle T_{on}\rangle}{\langle T_s\rangle}(1+a p_\mu) +\frac{2}{1+p_\mu}(1+ap_\mu)^2 -a^2p_\mu-1,
\end{align}
which is the same condition as obtained in \eref{mfpt-cond} in the main text.
In the limit of $\langle T_s\rangle\gg \langle T_f\rangle, \langle T_{on}\rangle$ we have $a\to 0$. Further assuming that $a p_\mu \to 0$ we have 
\begin{align}
    CV_s >\sqrt{2\mu -1}.
\end{align}
As in Eq. (11) of the main text.

\section{Exact result for the MFPT and error rate under gamma distributed stall times}
\begin{figure}[b]
    \centering
    \includegraphics[width=7cm]{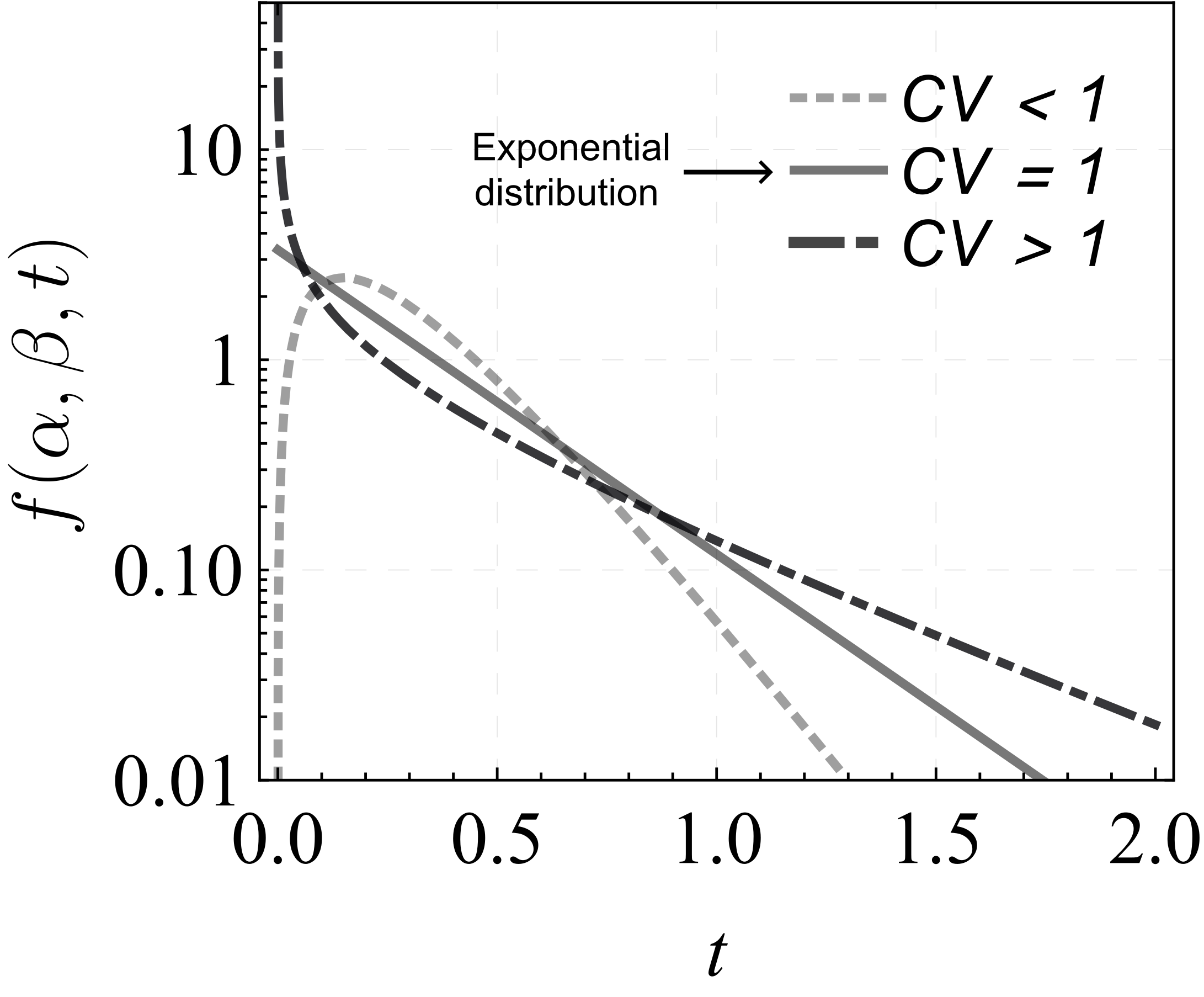}
    \caption{Variation of the Gamma distribution as in \eref{gam} for $\beta=\alpha/\tau_s$ with fixed mean $\tau_s=0.3$ but for different $CV=\frac{1}{\sqrt{\alpha}}$. For $\alpha=1$ or $CV=1$, we obtain the exponential distribution. Note that as $CV$ increases, the distribution starts to broaden, which physically implies that the role of fluctuation becomes crucial.}
    \label{figsm-1}
\end{figure}

In this section, we derive the exact results for the error rate and the MFPT for the replication process with gamma-distributed stall and forward waiting time. The gamma distribution of a random variable $T$ with rate $\beta$ and shape parameter $\alpha$ is given by
\begin{align}
   T\sim f(\alpha,\beta,t)=\frac{\beta^\alpha t^{\alpha-1}e^{-\beta t}}{\Gamma(\alpha)}. \label{gam}
\end{align}
The mean and co-efficient of variance of gamma distribution is found to be
\begin{align}
    \langle T \rangle =\frac{\alpha}{\beta},~~CV_T=\frac{1}{\sqrt{\alpha}}.
\end{align}
If we choose $\beta=\alpha/ \tau_s$ we have $\langle T \rangle=\tau_s$ so that varying $\alpha$ would only alter the $CV_T$ without changing the mean $\tau_s$. Interestingly, the variation of $\alpha$ drastically changes the shape of the distribution as well as shown in  \fref{figsm-1}.

For the purpose of illustrating the role of $CV$, we choose  
\begin{align}
    T_f \sim f(\alpha, \frac{\alpha}{\tau_f},t)~~~\text{and,}~~T_s \sim f(\alpha, \frac{\alpha}{\tau_s},t).\label{tfts}
\end{align}  
Throughout the article, we have always chosen $\tau_s>\tau_f=0.1$ such that the condition \eref{err-cond} is satisfied and the error is always reduced with proofreading. In addition, we choose $\langle T_{on}\rangle=0.01$ so that incorporation of a new subunit after excision is fast compared to the other system timescales. Furthermore, this time also ensures a physical condition that too much of the excision may hinder the assembly process as the overhead time $T_{on}$ accumulates. Plugging the results of \eref{tfts} in  \eref{err-exp} and \eref{mfpt-exp} we finally have the following results:
\begin{align}
   &\mu_{ec}= \frac{1}{1+10^{\alpha } p_{\mu } \left(\frac{\alpha +k_{ex} \tau _s}{10 \alpha +k_{ex}}\right)^{\alpha }}, ~~~ \langle T_{ec} \rangle=\frac{1}{100 k_{ex}} \left[ \frac{(k_{ex} + 100)(p_{\mu} + 1)}{ p_{\mu} \left( \frac{10\alpha}{10\alpha + k_{ex}} \right)^{\alpha} + \left( \frac{\alpha}{\alpha + k_{ex} \tau_s} \right)^{\alpha} } - 100 \right].
\end{align}
The above results for the error and MFPT have been plotted in \fref{fig2} in the main text.

\section{Derivation of conditions in \eref{cond-1} and \eref{cond-2} for T-cell response}
In this section, we exactly derive the condition (\eref{cond-1} and \eref{cond-2}) to improve both the speed and accuracy of T-cell activation. We start by deriving the expression for the mean of the activation time of a TCR as in \eref{tcell-mfpt}. Following the network of TCR activation as in \fref{fig1}\textbf{b}, one can write a renewal equation for the activation time $T_{act}$ as
\begin{equation}
T_{act} = T_{\text{on}} + \left\{
\begin{aligned}
    &\left. \begin{array}{ll} 
        T_N & \text{if } T_N < T_d^{ns} \\ 
        T_d^{ns} + T_{act}' & \text{if } T_d^{ns} \leq T_N 
    \end{array} \right\} \text{Pr}(\rho) \\
    &\left. \begin{array}{ll} 
        T_N & ~\text{ if } T_N < T_d^s \\ 
        T_d^s + T_{act}'' &~~ \text{if } T_d^s \leq T_N 
    \end{array} ~\right\} \text{Pr}(1-\rho),
\end{aligned},
\right.
\label{renewal-tcr}
\end{equation}
where $T_{act}',T_{act}''$ are independent and identically distributed copy of $T_{act}$. The quantity $T_{on}$ stands for the overhead time before a non-self antigen (with probability $\rho$) or a self antigen (with probability $1-\rho$) is attached to the TCR. The first two lines in \eref{renewal-tcr} denote the cases when a non-self antigen forms the complex C with the TCR. Upon formation of the complex, an intermediate cascaded reaction of $N$ steps is initiated, which ends at a random time $T_N$. If the TCR-pMHC complex is intact by then so that $T_N<T_d^{ns}$, the TCR is activated. On the contrary, if the complex dissociates before activation, such that $T_d^{ns}<T_N$, the time consumed is $T_d^{ns}$ after which the process restarts and the new activation time is given by $T_{act}'$. The same arguments can be extended to the self-antigens as well to explain the last two lines of \eref{renewal-tcr}. Finally, taking expectation over both sides of \eref{renewal-tcr} and after a slight rearrangement, we arrive at the result for the MFPT as
\begin{align}
     \langle T_{act}\rangle= \frac{\langle T_{on} \rangle + \rho \langle \text{min}(T_N, T_d^{ns}) \rangle + (1-\rho) \langle \text{min}(T_N, T_d^{s}) \rangle}{1-\rho \Pr (T_d^{ns}\leq T_N) -(1-\rho)\Pr (T_d^{s}\leq T_N)}. \label{tcr-mfpt-1}
\end{align}
Although the above results hold equally true for any arbitrary dwell time distribution of antigens on the TCR, we consider the paradigmatic choice of unbinding with a dissociation rate $k_{\text{off}}$ such that the dwell time is exponentially distributed. Particularly, we assume that the dissociation rate for the non-self antigen is $m$-fold slower than the self one, leading to a higher signalling probability. We thus have
\begin{align}
    T_d^{s}\sim k_{\text{off}}e^{- k_{\text{off}} t},~~ \text{and}~~~ T_d^{ns}\sim mk_{\text{off}}e^{- mk_{\text{off}} t}. \label{dt-dist}
\end{align}
Plugging these distributions into \eref{tcr-mfpt-1} we arrive at
\begin{align}
     \langle T_{act}\rangle= \frac{\langle T_{on} \rangle + \rho \left(\frac{1-\widetilde{T}_N(m k_\text{off})}{m k_\text{off}}\right) + (1-\rho) \left(\frac{1-\widetilde{T}_N( k_\text{off})}{k_\text{off}}\right)}{1-\rho \left(1-\widetilde{T}_N(m k_\text{off})\right) -(1-\rho)\left(1-\widetilde{T}_N( k_\text{off})\right)}. \label{tcr-mfpt-2}
\end{align}
To understand whether the TCR-pMHC unbinding actually improves the speed of the response, in a similar vein to the replication process, we again take the limit $k_\text{off}\to 0$ and consider terms only up to $\mathcal{O}(k_{\text{off}})$ to have
\begin{align}
     \langle T_{act}\rangle_{k_\text{off}\to 0}&= \frac{\langle T_{on} \rangle + \rho \left(\langle T_N \rangle -\frac{m k_\text{off} }{2}\langle T_N^2 \rangle\right) + (1-\rho) \left(\langle T_N \rangle -\frac{ k_\text{off} }{2}\langle T_N^2 \rangle\right)}{1-\rho m k_\text{off}\langle T_N \rangle -(1-\rho)k_\text{off}\langle T_N \rangle} \nonumber\\
     &=\left[\langle T_{on} \rangle +\langle T_N \rangle -\frac{k_\text{off}\langle T_N^2 \rangle}{2}\left(m\rho +(1-\rho)\right)\right]\times \left[1+k_\text{off}\langle T_N \rangle\left(m\rho +(1-\rho)\right)\right]\nonumber\\
     &=\langle T_{on} \rangle +\langle T_N \rangle + k_\text{off}\left(m\rho +(1-\rho)\right)\left(-\frac{\langle T_N ^2 \rangle}{2}+\langle T_{on}\rangle \langle T_N\rangle + \langle T_N\rangle^2\right).
\end{align}
For the mean activation time to reduce, we must have the contribution from the second term to be negative, such that
\begin{align}
  & -\frac{\langle T_N ^2 \rangle}{2}+\langle T_{on}\rangle \langle T_N\rangle + \langle T_N\rangle^2<0,\nonumber\\
   \implies& CV_{T_N}^2=\frac{\langle T_N ^2 \rangle-\langle T_N \rangle^2}{\langle T_N \rangle^2}>1+\frac{2 \langle T_{on} \rangle}{\langle T_N \rangle},
\end{align}
which is the same condition as presented in the main text (\eref{cond-1}).
Let us now shift our attention to the error rate of activation. We recall the expression for the error rate as in \eref{tcell-err}
\begin{align}
    \eta=1-\rho \text{Pr}(T_N<T_d^{ns}) -(1-\rho)\text{Pr}(T_d^{s}<T_N).
\end{align}
Plugging the distributions from \eref{dt-dist} in the above equaiton we have
\begin{align}
    \eta&=1-\rho\widetilde{T}_N(m k_\text{off})-(1-\rho)(1-\widetilde{T}_N( k_\text{off})).
\end{align}
Again taking the limit $k_\text{off}\to 0$ and consider terms only up to $\mathcal{O}(k_{\text{off}})$ we have
\begin{align}
    \eta_{k_{\text{off}} \to 0}&= 1 -\rho(1-m k_\text{off} \langle T_N\rangle)-(1-\rho)k_\text{off}\langle T_N\rangle\nonumber\\
    &= 1-\rho + k_\text{off} \langle T_N\rangle \left(\rho m -(1-\rho)\right).
\end{align}
Evidently, for error rate to decrease we must have 
\begin{align}
    &\rho m -(1-\rho)<0\nonumber\\
   \implies &m <\frac{1-\rho}{\rho}<1,
\end{align}
as shown in \eref{cond-2} of the main text.

\end{document}